\documentclass[12pt]{iopart}
\usepackage{graphicx}
\def\R{\partial}

\begin{document}

\title[The dynamic exponent of the Ising model on negatively curved surfaces]
{The dynamic exponent of the Ising model on negatively curved surfaces}

\author{Hiroyuki Shima and Yasunori Sakaniwa}

\address{Department of Applied Physics, Graduate School of Engineering,
Hokkaido University, Sapporo 060-8628, Japan}
\ead{shima@eng.hokudai.ac.jp}

\begin{abstract}
We investigate the dynamic critical exponent of the two-dimensional Ising model that is defined on a curved surface with constant negative curvature. By using the short-time relaxation method, we find the quantitative alteration of the dynamic exponent from the known value for the planar Ising model. This phenomenon is attributed to the fact that the Ising lattices embedded on negatively curved surfaces act as those in an infinite dimension, thus yielding the dynamic exponent deduced from the mean-field theory. We further demonstrate that the static critical exponent for the correlation length exhibits the mean-field exponent, which agrees with the existing results obtained by canonical Monte Carlo simulations.
\end{abstract}


\maketitle

\section{Introduction}
\label{sec_1}

Physics on curved surfaces is an intriguing subject especially 
with regard to second-order phase transition.
This is because the non-zero curvature of the underlying surface converts
the essential symmetries of the embedded physical system
through a spatial fluctuation of the metric.
The two-dimensional (2$d$) Ising model defined on a curved surface
is an exemplary system.
While the interest in this model originates from its relevance 
to the quantum gravity theory \cite{grav1,grav2,grav3,grav4,grav5},
it is presently motivated by the remarkable advances in nanotechnology 
that enables the production of curved magnetic layers of desired shapes 
\cite{mag1,mag2}.
Earlier studies have revealed that
the Ising model defined on a curved surface exhibits peculiar critical behaviors
that differ from those of the planar Ising model
\cite{sp1,sp2,sp3,sp4,sp6,sp7}.
In particular, on a surface with a constant negative curvature,
several static critical exponents have been proven to 
deviate quantitatively from the exact solution for the planar model
\cite{Shima}.
The latter results indicate the occurrence of a novel universality class of the $2d$
Ising model induced by surface curvature.

Successful findings on the static critical properties
motivate the study of the dynamic critical behaviors of the Ising model
on a curved surface.
The dynamic scaling hypothesis states that,
in the vicinity of the critical temperature $T_c$,
the thermodynamic quantities $O$ of finite-sized systems
obey the dynamic scaling form as follows \cite{Cardy,Pasq}:
\begin{equation}
O(t, \epsilon, L) = L^{\phi} \cdot {\cal O}\left( t L^{-z},\; \epsilon L^{1/\nu} \right),
\label{eq_01}
\end{equation}
where $t$ is the dynamic time variable,
$\epsilon=|T-T_c|/T_c$ the reduced temperature,
$L$ the linear dimension of the system,
and ${\cal O}$ the universal scaling function.
$\phi$ and $\nu$ are static critical exponents,
while $z$ is the dynamic critical exponent
that takes the value of $z\simeq 2.2$ in the planar Ising model
(See Refs.~\cite{Night1,Soares,Night2} and references therein).
The main concern is whether the finite surface curvature leads to a change 
in the value of $z$ given above.
It should be noted that the curvature-induced alteration of static critical exponents
observed in Ref.~\cite{Shima}
has no bearing on this problem,
since the value of $z$ that characterizes the dynamic universality class of the system
is generally independent of its static universality class.
Determination of $z$ is, therefore,
crucial to obtain a better understanding of the curvature effect
on the critical behavior of the $2d$ Ising model.

In the present study,
we numerically investigate the dynamic critical exponent
of the Ising model defined on a curved surface
with constant negative curvature.
The short-time relaxation (STR) method \cite{Soares,str1,Janssen,str2}
and the finite-size scaling analysis reveal that 
the dynamic exponent on the curved surface exhibits a value 
that is different from that for the planar Ising model.
This quantitative change in the dynamic exponent results from the peculiar intrinsic
geometry of the underlying surface on which Ising lattices act similar to those 
in an infinite dimension; therefore,
the dynamic exponent on negatively curved surfaces
yields the mean-field behavior in the thermodynamic limit.
In addition, we observe that the static critical exponent and the critical temperature
determined by the STR method also exhibit the mean-field behavior,
as has been demonstrated by canonical Monte Carlo (MC) simulations.
The latter finding supports the conclusion of our previous work \cite{Shima}.

\section{Ising lattices on a curved surface}
\label{sec_2}

In order to extract the curvature effect on the critical behavior of the Ising lattice model,
it is desirable to adopt a simply connected surface with a constant curvature.
Although a spherical surface is an optimal geometry,
it precludes from taking the thermodynamic limit
while maintaining its {\it positive} curvature,
since it reduces to a flat plane at this limit.
Instead, we focus on a curved surface with {\it negative} constant curvature
--- {\it a pseudosphere} \cite{Coxeter,Firby}.
The pseudosphere is a simply connected infinite surface
in which the Gaussian curvature at arbitrary points
possesses a constant negative value.
Hence, it is a suitable geometric surface
for the consideration of the curvature effect on the critical properties
of a system.
Although the Ising models embedded on a pseudosphere have been considered thus far
\cite{Rietman,Elser,Angles,Doyon},
their critical dynamics have yet to be explored.
The pseudosphere has also been considered with respect to various physical issues,
where its intrinsic geometry is relevant to the nature of the system.
These issues range from quantum physics \cite{Balazs,Avron}, the string theory \cite{string}
to cosmology \cite{Levin}.

It is intriguing that
the constant negative curvature of
a pseudosphere permits the establishment of a wide variety of regular lattices
\cite{Firby}.
The family of possible lattice structures consists of regular $p$-sided polygons
that satisfy the following inequality:
\begin{equation}
(p-2)(q-2) > 4,
\label{eq_02}
\end{equation}
where $q$ is the number of polygons that meet at each vertex.
The series of integer sets $\{p,q\}$ that satisfy (\ref{eq_02})
results in an infinite number of possible regular tessellations of a pseudosphere.
This is in contrast to the cases of a flat plane and a spherical surface,
where only a few kinds of regular tessellations are possible to realize.

\begin{figure}
\begin{center}
\includegraphics[width=5.0cm]{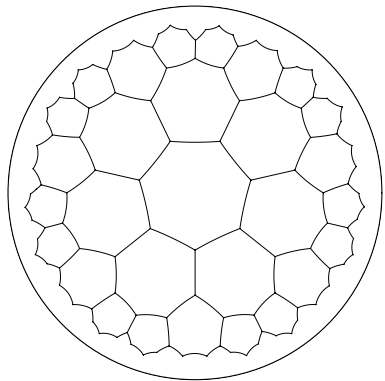}
\end{center}
\caption{Regular heptagonal lattice established on the Poincar\'e disk.
The number of concentric layers of heptagons is $L=3$ in this figure.
All heptagons depicted within the circle
are congruent with respect to the metric given in Eq.~(\ref{eq_03}).
The circumference corresponds to an infinite distance from the center of the circle.}
\label{fig_1}
\end{figure}

Amongst the infinite number of choices,
we focus on a heptagonal $\{7,3\}$ tessellation
in order to construct the Ising lattice on a pseudosphere.
The resulting lattice is depicted in Fig.~1 in terms of the Poincar\'e disk model,
which is a compact representation of a pseudosphere.
We emphasize that all the heptagons presented in Fig.~1
are congruent with respect to the metric on the disk that is expressed as follows:
\begin{equation}
ds^2 = \frac{4(dr^2 + r^2 d\theta^2)}{(1-r^2)^2},
\label{eq_03}
\end{equation}
in the polar coordinate system.
The metric (\ref{eq_03}) ensures that the Gaussian curvature is $\kappa=-1$
at arbitrary points on the disk (See Appendix).
In addition, it leads to the conclusion that the circumference of the unit circle 
displayed in Fig.~\ref{fig_1} corresponds to an infinite distance from the origin $r=0$.
Therefore, the regular heptagonal lattice formed in the Poincar\'e disk
can assume an arbitrary large system size,
thereby facilitating the measurement of physical quantities at the thermodynamic limit
while keeping curvature constant.

The size of our lattice is characterized by 
the number of concentric layers of heptagons, denoted by $L$,
which effectively functions as the linear dimension in our lattice.
It is noteworthy that the total number of sites $N$ for our heptagonal lattices
exhibit a complicated dependence on $L$ that is expressed as
\begin{eqnarray}
N(1) &=& 7, \nonumber \\
N(L) &=& 7+7 \sum_{j=0}^{L-2}
\left(
u_+ {v_+}^j + u_- {v_-}^j
\right),
\; [L\ge 2]
\label{eq_04a}
\end{eqnarray}
where $u_{\pm}$ and $v_{\pm}$ are defined by
\begin{equation}
u_{\pm} = 2 \pm \sqrt{5}, \;\;\; v_{\pm}=\frac{1+u_{\pm}}{2}.
\label{eq_03a}
\end{equation}
When $L\gg 1$, the total sites $N$ grows asymptotically as $N\propto e^L$,
which is quite rapid in comparison with the case of the planar Ising model:
$N\propto L^2$.
This exponential increase in $N$
is a manifestation of the constant negative curvature of the underlying geometry,
resulting in that the ratio $[N(L)-N(L-1)]/N(L)$ approaches
a non-zero constant $1-e^{-1}$ in the limit $L\to \infty$.
The latter feature means that the boundary sites of our lattices can not be neglected
even in the thermodynamic limit,
but contain a finite fraction of the total sites.

Boundary effects coming from these sites
are difficult to be eliminated completely,
because the periodic boundary conditions are hard to be employed to regular
lattices assigned on a pseudosphere.
Thereby, we have used the following manner in order to reduce the boundary effects
on physical quantities of the system.
Suppose that an Ising lattice consists of $L_{\rm out}$
concentric layers of heptagons.
Then, for employing the STR method (see Sec.~3 for its details),
we take into account only the Ising spins involved in
the interior $L_{\rm in}$ layers ($L_{\rm in} \le L_{\rm out}$)
so as to reduce the contribution of the spins locating near the boundary.
In actual calculations, $L_{\rm in}$ is varied from 4 to 6,
and for each $L_{\rm in}$ 
the number of disregarded layers 
$\delta L \equiv L_{\rm out}- L_{\rm in}$ is systematically increased from 0 to 3.
By examining the asymptotic behavior for large $\delta L$,
we can deduce the bulk properties of the Ising lattice model
embedded on the pseudosphere.
Similar procedure has been employed thus far \cite{Rietman,Doyon} to extract
 the bulk critical properties of the negatively curved system.

%
\section{Short-time relaxation method}
%

Our aim is to investigate the dynamic critical behavior
of the ferromagnetic Ising model defined on the heptagonal lattice.
The Hamiltonian of the system is given by
\begin{equation}
H = -J \sum_{<i,j>} s_i s_j,\quad s_i=\pm 1,
\label{eq_04}
\end{equation}
where $\langle i,j \rangle$ denotes a pair of nearest-neighbour sites
and $J(>0)$ is the coupling strength between them.
Hereafter, temperatures and energies are expressed in units of
$J/k_{\rm B}$ and $J$, respectively.

In order to determine the dynamic critical exponent,
we use the short-time relaxation (STR) method \cite{Soares,str1,Janssen,str2}.
This method enables neglecting equilibration,
thereby avoiding computational burden significantly
as compared to dynamical simulations in equilibrium.
The basis of the STR method is the measurement of the quantity \cite{Soares,Olive}
\begin{equation}
Q(t) = \left\langle {\rm sgn} 
\left( \sum_{i=1}^N s_i(t) \right) \right\rangle,
\;\;
{\rm sgn}(x)=\left\{
\begin{array}{rr}
1 & (x>0), \\
-1 & (x<0),
\end{array}
\right.
\label{eq_05}
\end{equation}
which is obtained by the conventional MC procedure.
Here, the time $t$ is measured in terms of one MC sweep,
and the angular bracket indicates that the average is taken over different time sequences
that begin from the same initial configuration.
In actual simulations,
the initial configuration is selected as $s_i=1$ for all sites,
thereby yielding $Q(0)=1$.
It then follows a monotonic decay with an increase in $t$,
and finally $Q(\infty)=0$,
since there is no preferred direction at equilibrium ($t\to \infty$).

A dynamic scaling hypothesis suggests that
in the vicinity of the critical temperature $T_c$,
the quantity $Q$ obeys the following scaling form:
\begin{equation}
Q(t,\epsilon,N) = {\cal Q}\left( t N^{-\bar{z}}, \epsilon N^{1/\mu} \right).
\label{eq_06}
\end{equation}
Here, $\bar{z}$ and $\mu$ are referred to as the dynamic and static critical exponent,
respectively, for the Ising lattices on the pseudosphere
(The physical meaning of these exponents will be given soon below).
It is noted that $N$ instead of $L$ is employed in (\ref{eq_06})
as a scaling variable;
this is a direct consequence of the exponential increase $N\propto e^L$
discussed in Sec.~2\footnote{Similar argument has been made regarding
an infinitely coordinated Ising model \cite{Nscale}
and a randomly-connected Ising model \cite{Das},
where $N$ is the only variable determining the system size.}.

The dynamic exponent $\bar{z}$ is determined as follows.
At the transition point $T=T_c$, $\epsilon=0$  so that
the second argument $\epsilon N^{1/\mu}$ of the scaling function ${\cal Q}$ vanishes.
This yields
\begin{equation}
\left.Q(t,N)\right|_{\epsilon=0} = f(x)
\label{eq_07}
\end{equation}
with the definition $x=t N^{-\bar{z}}$.
Equation (\ref{eq_07}) states that, if we calculate $Q(t)$ by maintaining $T=T_c$
and plot the results against $x$ by selecting an appropriate value of $\bar{z}$,
the curves of $Q(t)$ for different $N$s
should collapse onto a single curve.
Hence, the dynamical exponent $\bar{z}$ can be identified 
as the optimal fitting parameter for Eq.~(\ref{eq_07}).
We should note that the evaluation of $\bar{z}$ by the above procedure
requires the knowledge of the value of $T_c$ beforehand;
therefore, in actual calculations we use the numerical data of $T_c$
obtained from canonical Monte Carlo simulations for the same model \cite{Shima}.

Apart from the value of $\bar{z}$, 
the STR method enables us to determine the static critical exponent $\mu$.
This is achieved by extracting the data $Q(t_0)$
at the specific time $t=t_0$ for various $T$ and $N$.
The time $t_0$ is defined such that parameter $a=t_0 N^{-\bar{z}}$
with fixed $\bar{z}$
is invariant to the change in $N$.
Under this condition, we obtain
\begin{equation}
\left.Q(\epsilon,N)\right|_{t=t_0} = g(y),
\label{eq_08}
\end{equation}
where $y$ is defined as $y=\epsilon N^{1/\mu}$.
As a result, fitting the data of $Q(t_0)$ for different $T$ and $N$
onto a single curve against $y$
yields the values of $\mu$ and $T_c$ as the optimal fitting parameters.
This procedure provides a complementary method to determine $\mu$ and $T_c$,
which were extracted in \cite{Shima}.

It deserves further comment about the exponents $\bar{z}$ and $\mu$ 
introduced in (\ref{eq_06}).
From the scaling hypothesis,
the exponent $\bar{z}$ is assumed to describe the dynamic scaling law 
\begin{equation}
\tau\propto {\xi_V}^{\bar{z}}
\label{eq_08a}
\end{equation}
between the relaxation time $\tau$ and the correlation {\it volume} $\xi_V$.
Here, $\xi_V$ is a natural generalization \cite{Nscale,swn} 
of the localization length $\xi$,
thus obeying the power law $\xi_V \propto \epsilon^{-\mu}$ in the vicinity of $T_c$.
When the underlying geometry is a $d$-dimensional flat surface (or space),
$\xi_V \propto \xi^d$ holds so that (\ref{eq_08a}) 
yields $\bar{z}=z/d \sim 1.1$ for two-dimensional planar Ising models\footnote{In
$d$-dimensional planar Ising models,
$\xi_V \propto \xi^d$ so that $\tau\propto \xi^z = {\xi_V}^{z/d} = {\xi_V}^{\bar{z}}$,
and $\xi_V \propto \epsilon^{-\mu} = \epsilon^{-\nu d}$ since $\xi\propto \epsilon^{-\nu}$.
Eventually we obtain $\bar{z}=z/d$ and $\mu=\nu d$.}.
However, when the underlying geometry is curved,
the relation $\xi_V \propto \xi^d$ becomes invalid so that
$\bar{z}$ may take a value different from the above.
Similarly, the static exponent $\mu$ reduces to $\mu=\nu d = 2$
for planar Ising models (because $\nu=1$ \cite{nu}),
while it is not the case for curved cases.
In fact, our numerical simulations have revealed that
the dynamic exponent $\bar{z}$ for heptagonal Ising lattices
exhibits a quantitative deviation from that for planar ones,
as will be seen in the next section.

%
\section{Results: Dynamic exponent $\bar{z}$}
%

\subsection{$\bar{z}$ for entire lattices $(\delta L=0)$}

\begin{figure}[ttt]
\begin{center}
\includegraphics[width=5.1cm]{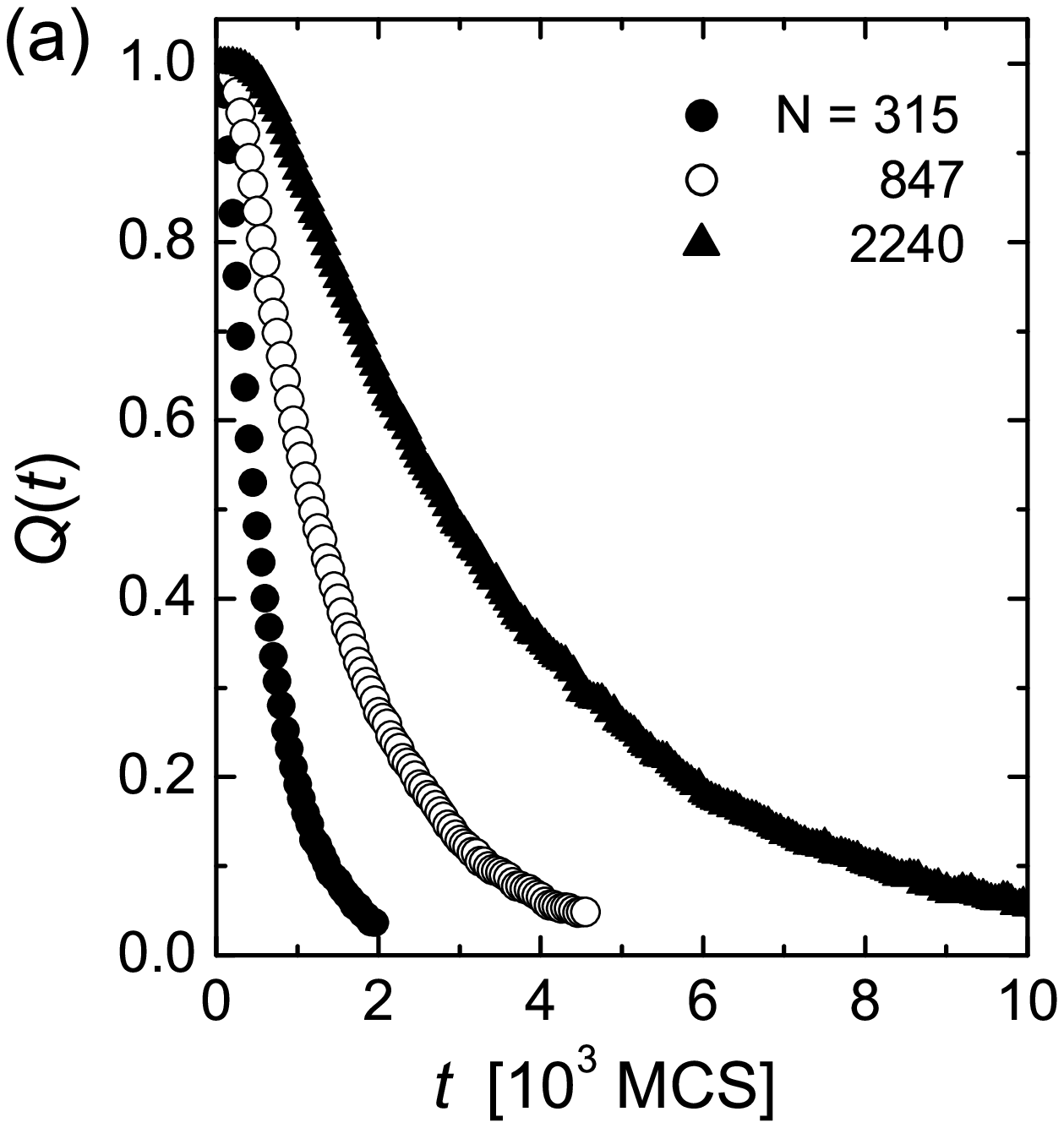}
\hspace*{-6mm}
\includegraphics[width=5.1cm]{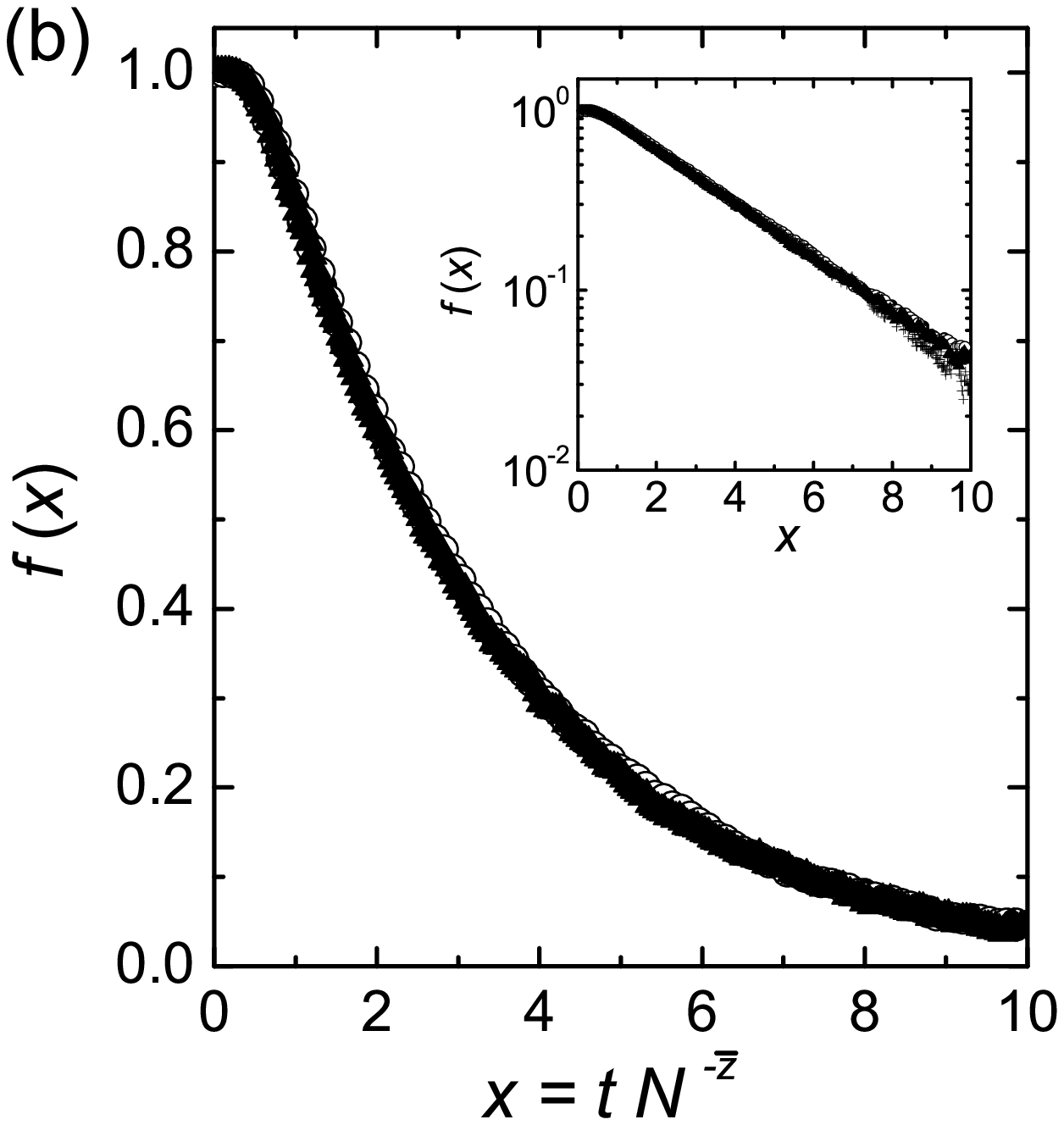}
\hspace*{3mm}
\includegraphics[width=5.1cm]{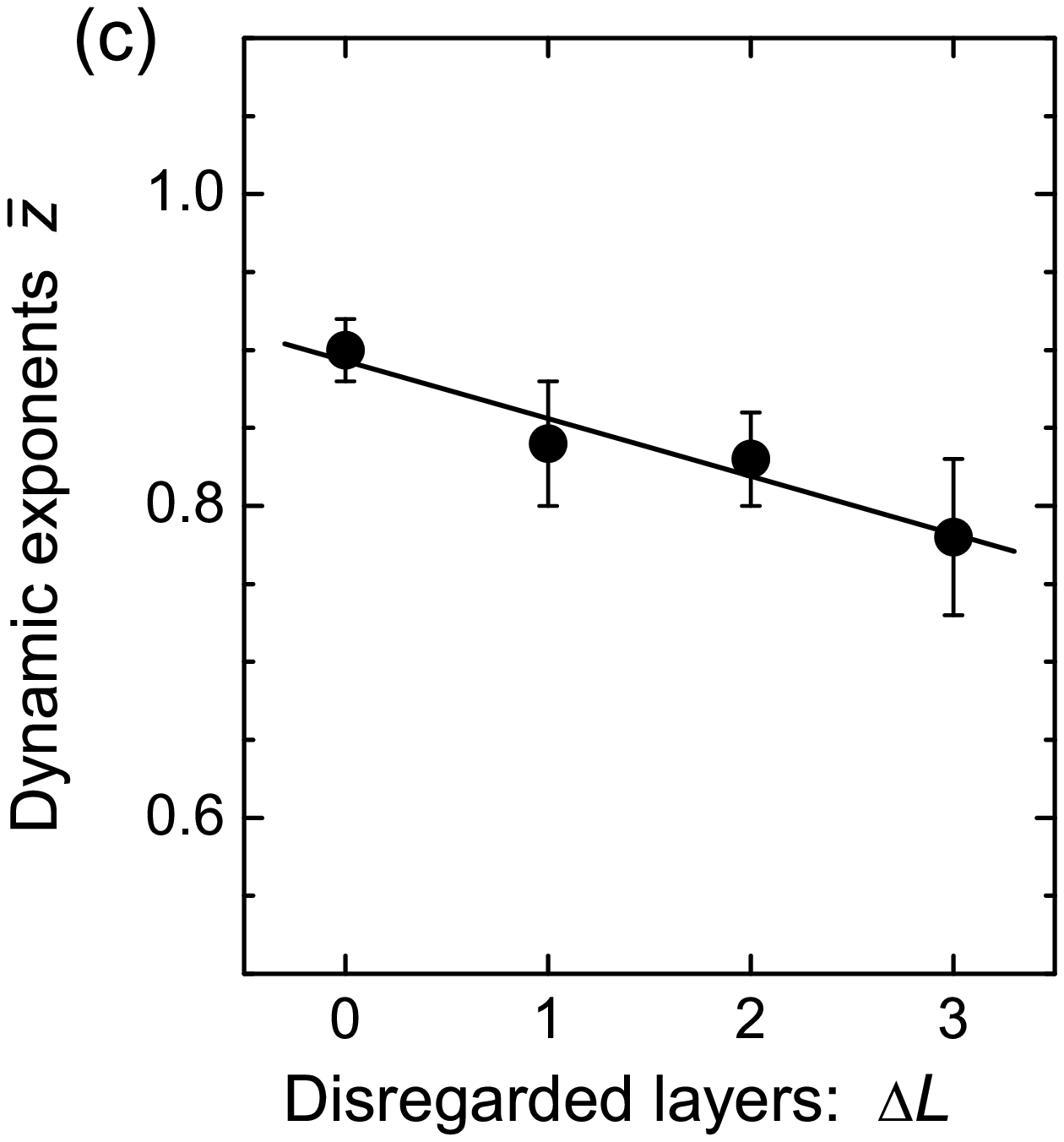}
\end{center}
\caption{(a) Calculated results of $Q(t)$ for various lattice sizes $N$.
(b) Finite size scaling of $Q(t,N)$ against the scaling variable 
$x=t N^{-\bar{z}}$.
The critical temperature $T_c=1.25$ is assumed in accordance with Ref.~\cite{Shima}.
The estimated value of $\bar{z}$ is $\bar{z}=0.90(2)$ with a 95\% confidence interval.
(c) The dependence of $\bar{z}$ on the number of disregarded layers $\delta L$.
The solid line serves as a guide to eye.}
\label{fig_2}
\end{figure}

Before addressing the bulk critical properties,
we first demonstrate the results for entire heptagonal lattices
with $\delta L = 0$ (boundary contributions are fully involved).
Figure \ref{fig_2} (a) shows the short-time relaxation behavior of $Q(t)$ 
for various systems sizes $N$.
The temperature is fixed at $T=T_c=1.25$ consistent with Ref.~\cite{Shima},
and the standard local update algorithm \cite{Binder} is employed 
in order to calculate the time evolution of $Q(t)$.
$N$ is varied from $315$ to $2240$;
this corresponds to the change in the number of concentric layers
from $L_{\rm in}=4$ to $L_{\rm in}=6$ (See Eq.~(\ref{eq_04a})).
For each value of $N$,
we have performed a sample average for more than $3\times 10^4$ independent runs.

We see that all curves of $Q(t)$ in Fig.~\ref{fig_2} (a) exhibit
a monotonic decay from the initial value of $Q(0)=1$
toward the equilibrium value $Q(\infty)=0$.
The decay time grows systematically
with an increase in the system size $N$
as expected from the scaling law $\tau\propto N^{\bar{z}}$.
Hence, rescaling the horizontal axis $t$ of Fig.~\ref{fig_2} (a) 
by dividing it by the factor $N^{\bar{z}}$ with an appropriate exponent $\bar{z}$
yields the fitting of all curves into a single curve.

Figure \ref{fig_2} (b) presents the scaling attempt of $Q$ 
by taking $x\equiv t N^{-\bar{z}}$ as a scaling argument.
The dynamic exponent $\bar{z}$ is determined as the optimal value
that minimizes the following quantity \cite{swn}:
\begin{equation}
\Delta(\bar{z}) 
= \sum_{i=1}^{n_{\rm seg}} \sum_{x\in X_i}
\left[ \frac{Q(x)-a_i -b_i x}{\sigma(x)} \right]^2,
\label{eq_09}
\end{equation}
where $a_i$ and $b_i$ are fitting parameters, and $\sigma(x)$ 
is the standard deviation of $Q$.
Note that in (\ref{eq_09}), 
the entire $x$ range is partitioned
into $n_{\rm seg}$ segments $X_i$ with equal intervals.
We have set $n_{\rm seg}=20$ in actual calculations,
and extracted the values of $a_i$ and $b_i$ $(1\le i \le 20)$
that minimize $\Delta(\bar{z})$ for a given $\bar{z}$.
By probing the behavior of $\Delta(\bar{z})$ with respect to $\bar{z}$,
we have attained the optimal value $\bar{z}=0.90(2)$
with a 95\% confidence interval.
The resulting plot shown in Fig.~\ref{fig_2} (b) displays the best collapse onto
a single curve $f(x)$ in a broad range of the scaling variable $x=t N^{-\bar{z}}$.
Importantly, the resulting value $\bar{z}\simeq 0.9$ is somewhat smaller than
that of the planar Ising model $\bar{z}=z/d \simeq 1.1$.
Since the boundary contribution is fully involved in the present case $(\delta L=0)$,
the above deviation of $\bar{z}$ possibly stems from the mixture effect
of the boundary Ising spins and the interior heptagonal Ising lattice.

\subsection{$\bar{z}$ for bulk lattices $(\delta L\ge 1)$}

We now turn to the study of bulk critical properties
of the heptagonal Ising lattice model.
As mentioned in Sec.~2, the boundary spins of Ising lattices on a pseudosphere
are thought to affect significantly to the nature of the system;
this is because the number of the spins along the boundary
increases as fast as that of total spins of the lattice.
Thereby, in order to extract the bulk critical exponents,
we try to remove the boundary contribution
by setting the disregarded layers 
$\delta L = L_{\rm out} - L_{\rm in}$ to be finite, i.e., 
by summing up only the spins within the interior $L_{\rm in}$ layers
when performing MC simulations
on the systems consisting of $L_{\rm out} (> L_{\rm in})$ layers.

The calculated results are summarized in Figure \ref{fig_2} (c);
each data of $\bar{z}$ was extracted by the scaling analysis
for lattices of $4\le L_{\rm in} \le 6$ and a given $\delta L$.
The appropriate value of $T_c$ for each $\delta L$ is again referred to
the results obtained from canonical MC simulations \cite{Shima}.
(Hence, the maximum system size we have treated reaches $L_{\rm out}=9$,
which corresponds to $N=40432$.)
We have observed that $\bar{z}$ monotonically decreases
with increasing $\delta L$ within the range $0\le \delta L \le 3$.
The monotonic decrease in $\bar{z}$ indicates that for sufficiently large $\delta L$,
$\bar{z}$ takes a value that is considerably smaller than the corresponding value
of the planar Ising models: $\bar{z}\simeq 1.1$.
Thus, it follows that the bulk Ising lattices assigned on the negatively curved surfaces
belong to a dynamic universality class distinct from those of the planar Ising models.
The determination of the asymptotic value of $\bar{z}$ for $\delta L\gg 1$
would provide further conclusive information;
this would require a huge computational effort.
Instead, we have provided a concise argument regarding the asymptotic
value of $\bar{z}$ in Section 6;
this argument implies that the dynamic exponent in our system
takes the mean-field exponent $\bar{z}_{\rm MF} = 1/2$ in the limit $\delta L \to \infty$.

%
\section{Results: Static critical exponent $\mu$}
%

\begin{figure}
\begin{center}
\includegraphics[width=4.9cm]{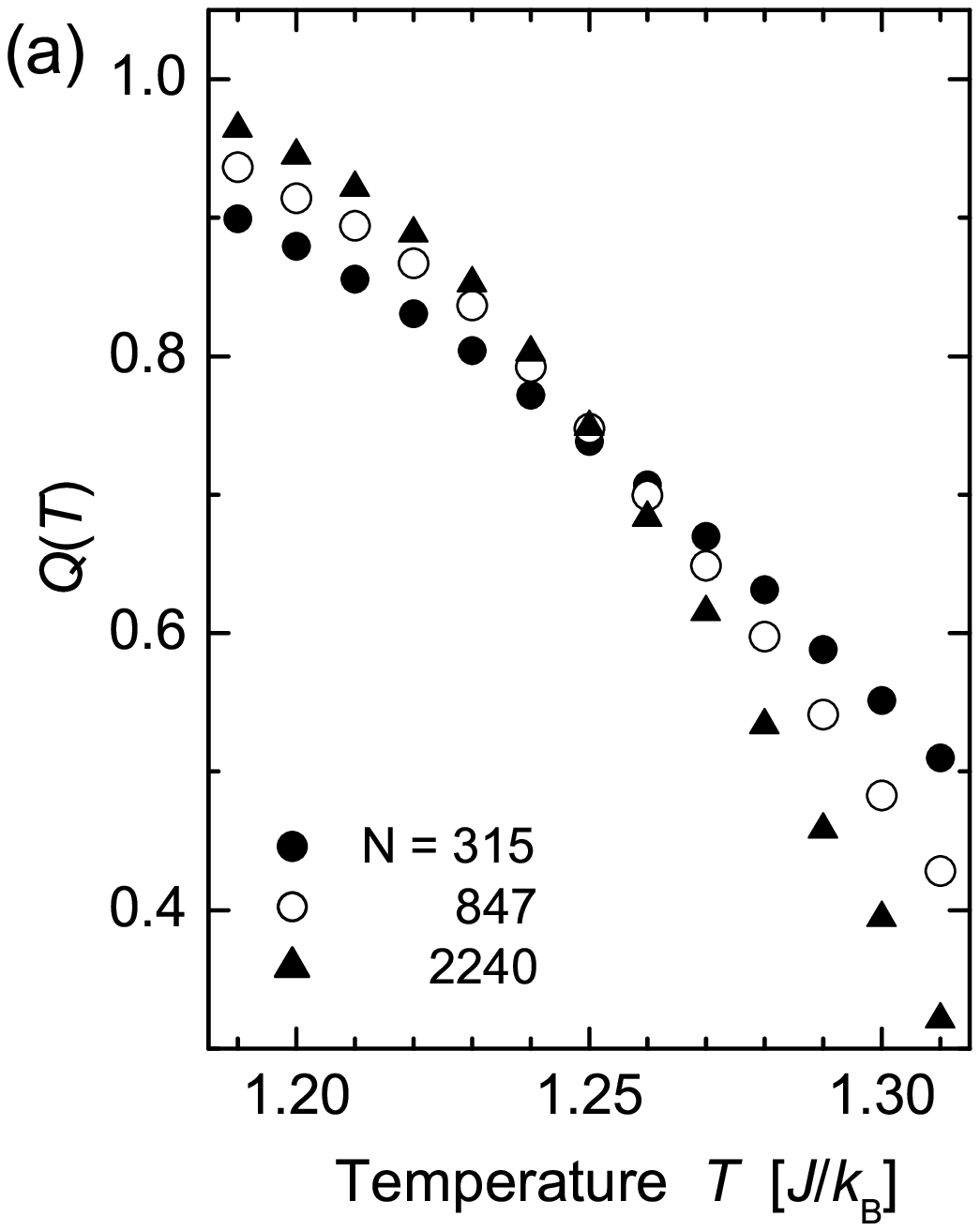}
\hspace{-6mm}
\includegraphics[width=4.9cm]{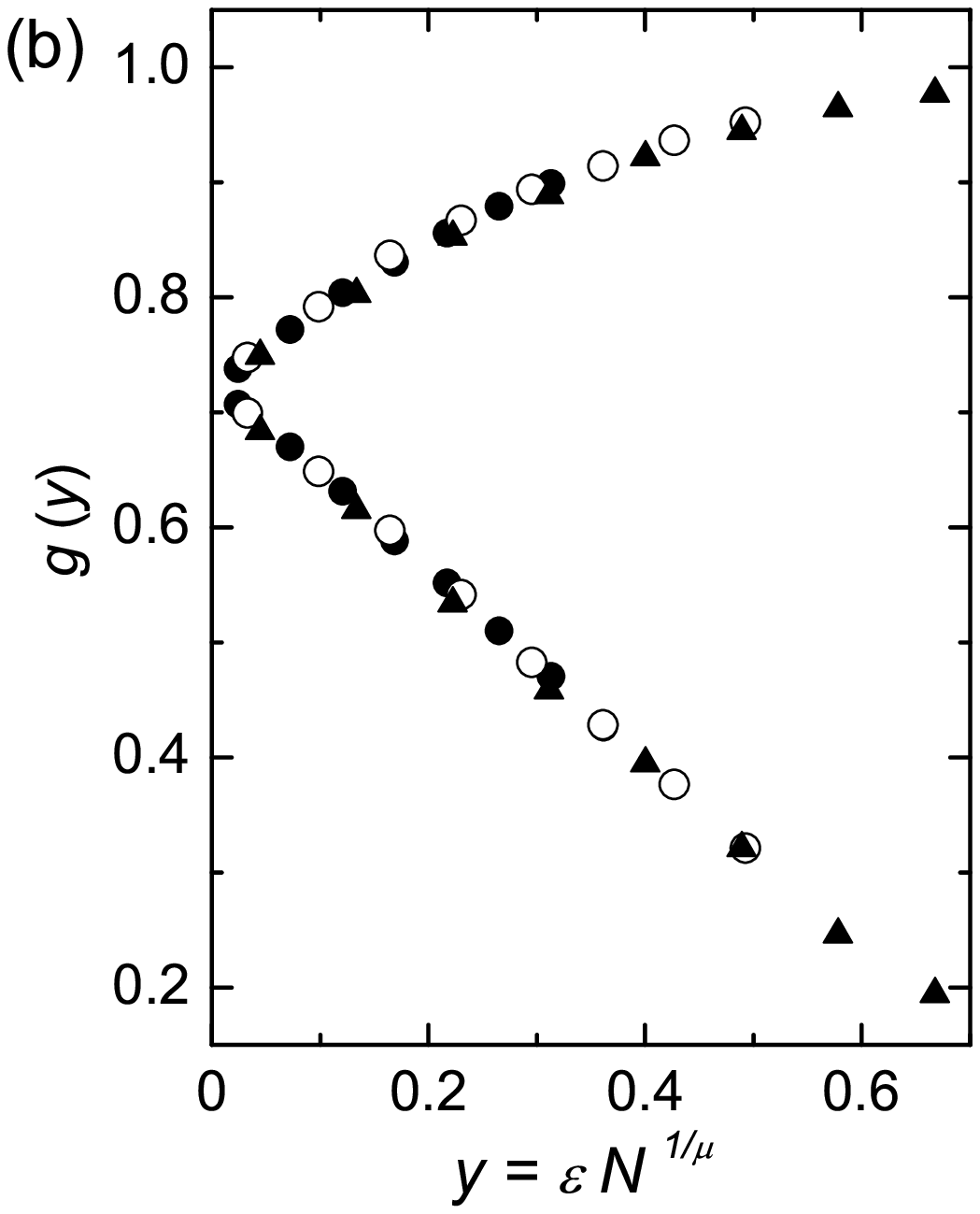}
\hspace{3mm}
\includegraphics[width=5.6cm]{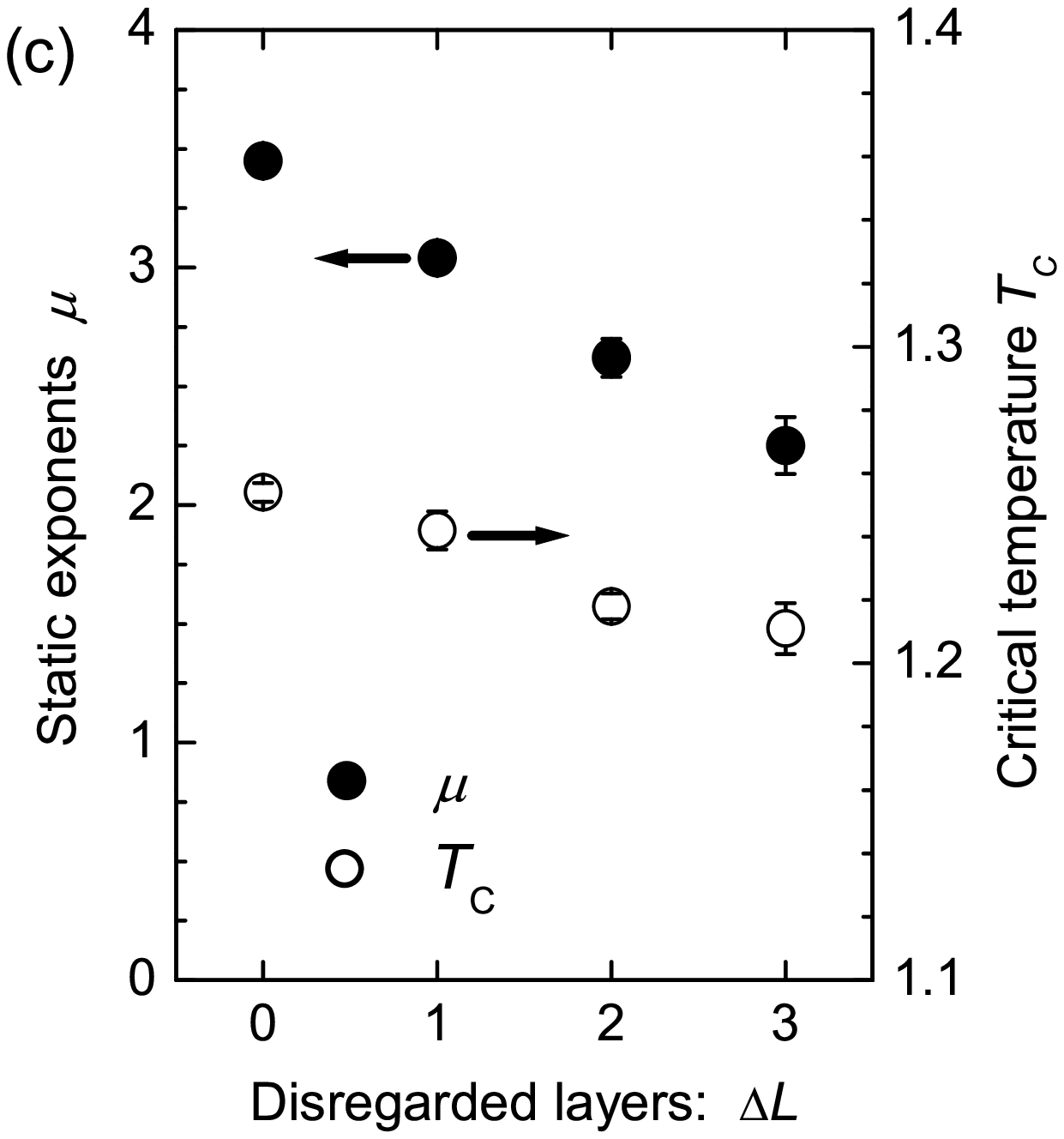}
\end{center}
\vspace*{-5mm}
\caption{(a) Calculated results of $Q(T)$ for various values of $N$.
Parameter $a=t N^{-\bar{z}}$ and the exponent $\bar{z}$ are 
fixed at $a=1.5$ and $\bar{z}=0.90$ for all values of $N$, respectively.
(b) A scaling plot of $Q(T,N)$ against the scaling variable $y=\epsilon N^{1/\mu}$.
The optimal values of $\mu$ for upper and lower branch are 
$\mu=3.45(4)$ and $\mu=3.36(3)$, respectively.
The critical temperature is estimated as $T_c=1.254(3)$ for the two branches.
(c) $\delta L$-dependences of $\mu$ and $T_c$.}
\label{fig_3}
\end{figure}

We now focus on the estimation of the static critical exponent $\mu$
and the critical temperature $T_c$ on the basis of the scaling relation (\ref{eq_08}).
Figure \ref{fig_3} (a) presents the $T$-dependence of $Q(T)$ 
under the condition $\delta L=0$.
Parameter $a=t N^{-z}$ is fixed at $a=1.5$,
where $\bar{z}=0.90$ is set in accordance with the result from Fig.~2 (b).
We observe a unique crossing point at $Q\simeq 0.74$ and $T\simeq 1.25$.
Even when the values of parameter $a$ are varied,
a crossing point still appears at a temperature almost identical to $T\simeq 1.25$,
as expected from Eq.~(\ref{eq_06}).
Hence, it provides a rough estimate of the critical temperature $T_c$.

The full scaling plot for $Q(T)$ is displayed in Fig.~\ref{fig_3} (b).
The fitting process was based on 
a polynomial expansion of the scaling function $g(y)$ expressed 
in Eq.~(\ref{eq_08}) as
\begin{equation}
g(y) \simeq \sum_{j=0}^{3} c_j \left( \epsilon N^{1/\mu} \right)^j,
\label{eq_10}
\end{equation}
where $c_j$, $T_c$, and $\mu$ are considered as fitting parameters.
A very smooth collapse is obtained with
$\mu=3.45(4)$ for the upper branch,
and $\mu=3.36(3)$ for the lower one.
The critical temperature is estimated as $T_c=1.254(3)$ for both branches.
It was also confirmed that the values of $\mu$ and $T_c$
are almost invariant to the change in the value of parameter $a$
that is arbitrarily chosen within the range of $0.5\le a \le 3.0$.

We have then clarified the boundary contribution to the values of $\mu$ and $T_c$
by setting $\delta L\ge 1$.
Figure \ref{fig_3} (c) shows the $\delta L$-dependence of $\nu$ and $T_c$
under the same numerical conditions with respect to $L_{\rm in}$ and $a$
as in the case of $\delta L=0$.
It is found that both $\mu$ and $T_c$ decreases monotonically 
with increasing $\delta L$.
Further noteworthy is the fact that all the values of $\mu$ and $T_c$
given in Fig.~\ref{fig_3} (c) are in excellent agreement with those
obtained by MC simulations in equilibrium \cite{Shima},
where the tendency of $\mu$ toward the mean field exponent $\mu_{\rm MF}=2$
in the limit $\delta L\to \infty$ has been suggested.
This consistency with respect to $\mu$ and $T_c$
strongly supports the validity of our STR approach to 
the critical behavior of the Ising model on curved surfaces.

%
\section{Discussions}
%

It has been demonstrated that the dynamic critical exponent
$\bar{z}$ of the heptagonal Ising lattice model exhibit a quantitative deviation from
that of the planar Ising model $\bar{z}\simeq 1.1$.
In fact, it takes a value $\simeq 0.9$ under the condition $\delta L=0$,
and still continues to decrease almost linearly with $\delta L$.
This behavior of $\bar{z}$ indicates the occurrence of
a novel universality class with regard to dynamic criticality,
and evidences the conjecture that the peculiar intrinsic geometry
of the underlying surface affects
the dynamic critical properties of the embedded system.
Besides, the static critical exponent $\mu$ also turns out to 
exhibit a monotonic decrease with increasing
$\delta L$ toward a particular value of $\mu_{\rm MF}=2$,
which is consistent with the value derived from the mean field theory.

The mean field behavior of $\mu$ has been suggested earlier
by Rietman {\it et al.} \cite{Rietman} and by Doyon and Fonseca \cite{Doyon}.
This behavior is attributed to the fact that
an Ising lattice embedded on a pseudosphere is effectively an infinite dimensional lattice
at large distance due to the exponential growth of the total spins $N$ \cite{Callan}.
For an ordinary Ising lattice in $d$ dimension,
the number of spins along the boundary, $N_s$, is expressed by $N_s \propto N^{1-(1/d)}$.
Hence, by comparing it with the peculiar relation $N_s \propto N$ that is valid 
on negatively curved surfaces,
we eventually reach the consequence $d= \infty$ for the latter geometry.
As a result, our heptagonal Ising lattices yield the mean-field critical exponent 
$\mu_{\rm MF} = \nu_{\rm MF} d_c = 2$ \cite{Nscale},
where $\nu_{\rm MF}=1/2$ is the mean-field exponent
and $d_c=4$ is the upper critical dimension for the Ising model.

In this context, the dynamic critical exponent $\bar{z}$
on negatively curved systems is supposed to exhibit the mean field exponent
$\bar{z}_{\rm MF}=z_{\rm MF}/d_c=1/2$ at $\delta L \gg 1$,
since $z_{\rm MF}=2$ for the Ising model.
This inference is indeed consistent with our numerical results; 
as shown in Fig.~2 (c), 
$\bar{z}$ takes a smaller value than the planar case ($\bar{z}\sim 1.1$) 
whatever value $\delta L$ is, and moreover, 
it monotonically decreases with reducing the boundary contribution up to $\delta L=3$.
Hence, when extrapolated to the bulk limit $\delta L \gg 1$,
it yields the mean field exponent $\bar{z}_{\rm MF}=1/2$ 
as similar to the case of static critical exponents \cite{Shima}.
We also note that it is interesting to investigate the dynamics exponents
associated with the Griffiths-type phase transition in the Ising model
defined on negatively curved surfaces \cite{Angles};
a comparison between the dynamics exponent of the bulk region 
and that of the boundary region will provide a better understanding
of the critical dynamics of entire Ising lattices 
defined on negatively curved surfaces.

Further noteworthy is that the surface curvature effect
on critical properties of the assigned spin lattice model 
would become richer if the spin variables 
possess orientational degrees of freedom.
In the latter cases,
the interacting energy of neighboring spins is determined by 
their relative angle,
which is a function of a spatially-dependent metric tensor.
Thereby, the Hamiltonian of the system involves the metric tensor
in an explicit form.
As a result,
the energetically preferable spin configurations
and the temporal fluctuations of the order parameters
are directly influenced by the
intrinsic geometry of the surface determined by the metric tensor.
It is thus expected that these vector-spin models 
defined on curved surfaces exhibit peculiarities
associated with phase transition and low energy excitations
\cite{Hei1,Hei2,Hei3,Hei4,Hei5}.
To the best of our knowledge,
there was no attempt to reveal the dynamic critical exponent
of vector-spin lattice models defined on curved surfaces.
The STR method that we have utilized is applicable to these systems;
detailed analyses of this issue will be presented in future.

%
\section{Conclusion}
%

In the present study,
we have investigated the dynamic critical behavior of the regular heptagonal Ising model 
defined on a curved surface with constant negative curvature.
The STR methods that incorporate finite-size scaling analyses
have been employed in order to compute the dynamical critical exponent $\bar{z}$
as well as the static critical exponent $\mu$ for the correlation volume $\xi_V$.
The resulting values of both the exponents are distinct from those of the planar Ising model.
Furthermore, we have studied quantitatively how the boundary spins contribute
to the determination of $\bar{z}$ and $\mu$,
and eventually we have revealed that they both reduce to the mean field exponent 
$\bar{z}_{\rm MF}=1/2$ and $\mu_{\rm MF} = 2$ in the bulk limit.
We hope that our results prove to be a fundamental basis for further fruitful studies
on the critical behaviors occurring on curved surfaces.

\ack

We thank T.~Nakayama and K.~Yakubo for fruitful discussions.
This work was supported in part by a Grant--in--Aid for Scientific
Research from the Japan Ministry of Education, Science, Sports and Culture.
One of the authors (H.S) thanks the financial supports from 
the 21st Century COE ``Topological Sciences and Technologies".

\clearpage
\appendix
\section{The Gaussian curvature in the Poincar\'e disk}

This appendix demonstrates that the Poincar\'e disk
possesses the negative constant value $\kappa=-1$
at an arbitrary point within the disk.
The Gaussian curvature $\kappa$ at a given position is defined by \cite{Landau}
\begin{equation}
\kappa \equiv -R/2, \label{eq_a01}
\end{equation}
with the quantities:
\begin{equation}
R \equiv g^{ij} R_{ij}, \;\;
R_{ij} \equiv g^{kl} R_{kilj}. \label{eq_a02}
\end{equation}
Here, $R$, $R_{ij}$, and $R_{kilj}$ are termed as the scalar curvature,
Ricci tensor, and curvature tensor, respectively.
(Hereafter, we will use the summation convention on repeated indices.)
The $g^{ij}$ is the inverse of the metric tensor $g_{ij}$ 
that determines the line element at a given position as
$ds^2 = g_{ij} dx^i dx^j$.
Note that all the three quantities $R$, $R_{ij}$, and $R_{kilj}$
are functions of $g_{ij}$.
Hence, in principle,
the knowledge of $g_{ij}$ and the explicit form of $R_{kilj}$
are sufficient to calculate the value of $\kappa$,
while the actual calculations are lengthy as shown below.

The explicit dependence of the curvature tensor $R_{kilj}$
on $g_{ij}$ is expressed as:
\begin{eqnarray}
R_{kilj} &=& g_{km} R^m_{\;\; ilj}, \label{eq_a04} \\
R^m_{\;\; ilj} &=&
 \R_l {\it \Gamma}^m_{\;\; ij} - \R_j {\it \Gamma}^m_{\;\; il}
+ {\it \Gamma}^m_{\;\; pl} {\it \Gamma}^p_{\;\; ij}
- {\it \Gamma}^m_{\;\; pj} {\it \Gamma}^p_{\;\; il},\;\;\;
\end{eqnarray}
where ${\it \Gamma}^m_{\;\; ij}$ is the connection defined by
\begin{equation}
{\it \Gamma}^m_{\;\; ij}
= \frac{g^{m p}}{2}
 \left( \R_j g_{p i} 
      + \R_i g_{p j} - \R_{p} g_{i j} \right).
\label{eq_a06}
\end{equation}
Equations (\ref{eq_a04})--(\ref{eq_a06}) yield
an alternative form of $R_{kilj}$,
which is expressed as:
\begin{eqnarray}
R_{kilj} &=&
\frac12 \left( \R_i \R_j g_{kl} + \R_k \R_l g_{ij} - \R_k \R_j g_{il} - \R_i \R_l g_{kj} \right)
\nonumber \\
& & +\; g_{rs} \left( {\it \Gamma}^r_{\; lk} {\it \Gamma}^s_{\; ij} 
                 - {\it \Gamma}^r_{\; jk} {\it \Gamma}^s_{\; il} \right).
\label{eq_a07}
\end{eqnarray}
It is noteworthy that
due to the symmetric property of Eq.~(\ref{eq_a07}),
the tensor $R_{kilj}$ has only four non-zero components in a two dimensional system.
Further, these components are related to each other in the following manner:
\begin{equation}
R_{1212} = \;- R_{2112} = \;-R_{1221} = R_{2121}.
\label{eq_a08}
\end{equation}
As a result,
the tensor $R_{kilj}$ is determined by only a single component $R_{1212}$.
In addition,
the symmetric property of $R_{kilj}$ results in the following equality
with respect to the components of the Ricci tensor $R_{ij}$:
\begin{equation}
R_{11} = g^{22} R_{2121}, \;\; R_{22} = g^{11} R_{1212},
\end{equation}
and $R_{12}=R_{21}=0$.
Consequently, we can represent the scalar curvature $R$ in two-dimensional systems
as
\begin{eqnarray}
R &=& g^{11} R_{11} + g^{22} R_{22} \nonumber \\ [2mm]
&=& g^{11}g^{22} R_{2121} + g^{22} g^{11} R_{1212} \nonumber \\ [2mm]
&=& \frac{2 R_{1212}}{g},
\label{eq_a20}
\end{eqnarray}
where $g= {\rm det} [g_{ij}]$, and thus $g^{11} = g_{22}/g$ and $g^{22} = g_{11}/g$.
Equation (\ref{eq_a20}) yields the key relation between the component $R_{2121}$
and the Gaussian curvature $\kappa$ as follows:
\begin{equation}
\kappa = -\frac{R_{1212}}{g}.
\label{eq_a25}
\end{equation}
This identity is known to hold for general two-dimensional systems \cite{Landau}.

The value of $R_{1212}$ for our model is evaluated by the following procedure.
For the Poincar\'e disk, the metric tensor $g_{ij}$
is represented in the matrix form as follows:
\begin{equation}
\left[ g_{ij} \right]= \left(
\begin{array}{cc}
4f & \;0 \\ [2mm]
0 & \;4r^2 f
\end{array}
\right),
\;\;\; f\equiv \frac{1}{(1-r^2)^2}.
\label{eq_a09}
\end{equation}
This means that the line element on the Poincar\'e disk is given by
$ds^2=4f (dr^2 + r^2 d\theta^2)$,
as already shown in Eq.~(\ref{eq_03}).
Straightforward calculation based on the metric (\ref{eq_a09})
reveals that there are four non-zero components of
the Christoffel symbol ${\it \Gamma}^m_{\;\; ij}$ expressed as
\begin{eqnarray}
{\it \Gamma}^1_{\;\; 11} &=& \frac{g^{11}}{2} \R_1 g_{11}, \quad 
{\it \Gamma}^1_{\;\; 22} = -\frac{g^{11}}{2} \R_1 g_{22}, \nonumber \\ [2mm]
{\it \Gamma}^2_{\;\; 12} &=& {\it \Gamma}^2_{\;\; 21} =  \frac{g^{22}}{2} \R_1 g_{22}.
\label{eq_a10}
\end{eqnarray}
By substituting Eqs.~(\ref{eq_a09}) and (\ref{eq_a10}) into 
the expression of the tensor $R_{kilj}$ (\ref{eq_a07}),
we directly obtain:
\begin{equation}
R_{1212} = \frac{16 r^2}{(1-r^2)^4} \; = \; \frac{1}{g}.
\end{equation}
Eventually, by comparing it with Eq.~(\ref{eq_a25}),
we arrive at the final conclusion
$\kappa = -1$ for arbitrary positions within the Poincar\'e disk.

%
%


\section*{References}

\begin{thebibliography}{99}
\bibitem{grav1}
Kazakov V A 1986 {\it Phys. Lett. A} {\bf 119} 140
\bibitem{grav2}
Crnkovic C, Ginparg P and Moore G 1990 {\it Phys. Lett. B} {\bf 237} 196
\bibitem{grav3}
Gross M and Harber H W 1991 {\it Nucl. Phys. B} {\bf 364} 703
\bibitem{grav4}
Francesco P Di, Ginsparg P and Zinn-Justin J 1995 {\it Phys. Rep.} {\bf 254} 1
\bibitem{grav5}
Holm C and Janke W 1996 {\it Phys. Lett. B} {\bf 375} 69
\bibitem{mag1} Himpsel F J, Ortega J E, Mankey G J and Willis R F 1998
{\it Adv.~Phys.} {\bf 47} 511
\bibitem{mag2} Mart\'in J I, Nogu\'es J, Liu K, Vicente J L,
and Schuller I K 2003 {\it J.~Magn.~Magn.~Mater.} {\bf 256} 449
\bibitem{sp1} Diego O, Gonzalez J and Salas J 1994 {\it J. Phys.} A: Math. Gen. {\bf 27} 2965
\bibitem{sp2} Hoelbling Ch and Lang C B 1996 {\it Phys. Rev.} B {\bf 54} 3434
\bibitem{sp3} Gonz\'alez J 2000 {\it Phys. Rev.} E {\bf 61} 3384
\bibitem{sp4} Weigel M and Janke W 2000 {\it Europhys. Lett.} {\bf 51} 578
\bibitem{sp6} Costa-Santos R 2003 {\it Phys. Rev. B} {\bf 68} 224423;
2003 {\it Acta. Phys. Pol. B} {\bf 34} 4777
\bibitem{sp7} Pleimling M 2004 {\it J. Phys. A} {\bf 37} R79
\bibitem{Shima} Shima H and Sakaniwa Y 2006 {\it J. Phys. A} {\bf 39} 4921
\bibitem{Cardy} Cardy J, 1996 {\it Scaling and Renormalization in Satistical Physics}
(Cambridge University Press, Cambridge UK)
\bibitem{Pasq} Calabrese P and Gambassi A 2005
{\it J. Phys. A} {\bf 38} R133
\bibitem{Night1} Nightingale M P and Bl\"ote H W J 1996 {\it Phys. Rev. Lett.} {\bf 76} 4548 
\bibitem{Soares} Soares M S, daSilva J K L and Barreto F C S 1997
{\it Phys. Rev. B} {\bf 55} 1021
\bibitem{Night2} Nightingale M P and Bl\"ote H W J 2000 {\it Phys. Rev. B} {\bf 62} 1089
\bibitem{str1} Li Z B, Sch\"ulke L and Zhang B 1995 {\it Phys. Rev. Lett.} {\bf 74} 3396
\bibitem{Janssen} Janssen H K, Schaub B and Schmittmann B 1989
{\it Z. Phys. B} {\bf 73} 539
\bibitem{str2} Zheng B 1998 {\it Int. J. Mod. Phys. B} {\bf 12} 1419
\bibitem{Coxeter} Coxeter H S M 1969 {\it Introduction to Geometry} (Wiley, New York)
\bibitem{Firby} Firby P A and Gardiner C F 1991
{\it Surface Topology} (Ellis Horwood, London)
\bibitem{Rietman} Rietman R, Nienhuis B and Oitmaa J 1992
{\it J. Phys.} A: Math. Gen. {\bf 25} 6577
\bibitem{Elser} Elser V and Zeng C 1993 {\it Phys. Rev.} B {\bf 48} 13647
\bibitem{Angles} d'Auriac J C A, M\'elin R, Chandra P and Dou\c{c}ot B 2001
{\it J. Phys.} A: Math. Gen. {\bf 34} 675
\bibitem{Doyon} Doyon B and Fonseca P 2004
{\it J. Stat. Mech.} P07002
\bibitem{Balazs} Balazs N L and Voros A 1986 {\it Phys.~Rep.} {\bf 143} 109
\bibitem{Avron} Avron J E, Klein M, Pnueli A and Sadun L 1992 {\it Phys. Rev. Lett.} {\bf 69} 128
\bibitem{string} D'Hoker E and Phong D H 1988 {\it Rev. Mod. Phys.} {\bf 60} 917
\bibitem{Levin} Levin J 2002 {\it Phys.~Rep.} {\bf 365} 251
\bibitem{Nscale} Botet R, Jullien R and Pfeuty P 1982 {\it Phys. Rev. Lett.} {\bf 49} 478
\bibitem{Das} Das P K and Sen P 2005 {\it Eur. Phys. J. B} {\bf 47} 391
\bibitem{Olive} deOliveira P M C 1992 {\it Europhys. Lett.} {\bf 20} 621
\bibitem{swn} Medvedyeva K, Holme P, Minnhagen P, and Kim B J 2003
{\it Phys. Rev. E} {\bf 67} 036118
\bibitem{Binder} Binder K and Heermann D W 2002
{\it Monte Carlo simulation in Statistical Physics}, (Springer)
\bibitem{nu} Kaufman B and Onsager L 1949 {\it Phys. Rev.} {\bf 76} 1244
\bibitem{Callan} Callan C and Wilczek F 1990 {\it Nucl. Phys.} B {\bf 340} 366
\bibitem{Hei1} Saxena A and Dandoloff R 1997 {\it Phys. Rev.} B {\bf 55} 11049
\bibitem{Hei2} Balakrishnan R and Saxena A 1998 {\it Phys. Rev.} B {\bf 58} 14383
\bibitem{Hei3} Freitas W A, Moura-Melo W A and Pereira A R 2005 {\it Phys. Lett.} A
{\bf 336} 412
\bibitem{Hei4} Pereira A R 2005 {\it J. Magn. Magn. Mater.} {\bf 285} 60
\bibitem{Hei5} Moura-Melo W A, Pereira A R, M\'ol L A S and Pires A S T 2006 {\it Phys. Lett.} A
{\it in press}; {\ttfamily cond-mat/0511443}
\bibitem{Landau} Landau L D and Lifshitz E M 1980 {\it The Classical Theory of Fields}
(Butterworth-Heinemann)




\end{thebibliography}
\end{document}